\documentclass[pra,twocolumn,preprintnumbers,superscriptaddress,longbibliography,showkeys]{revtex4-1}
\usepackage{amssymb}
\usepackage{amsmath,bm}
\usepackage{amsfonts}
\usepackage{wasysym}
\usepackage{graphicx}
\usepackage{dcolumn}
\usepackage{bm}
\usepackage{color}

\usepackage[colorlinks,linkcolor=blue, urlcolor=blue, anchorcolor=blue, citecolor=blue]{hyperref}



\begin{document}

\title{Fluctuation assisted collapses of Bose-Einstein condensates}
\date{\today}

\author{Junqiao Pan}
\affiliation{CAS Key Laboratory of Theoretical Physics, Institute of Theoretical Physics, Chinese Academy of Sciences, Beijing 100190, China}
\affiliation{School of Physical Sciences, University of Chinese Academy of Sciences, Beijing 100049, China}

\author{Yuqi Wang}
\affiliation{CAS Key Laboratory of Theoretical Physics, Institute of Theoretical Physics, Chinese Academy of Sciences, Beijing 100190, China}
\affiliation{School of Physical Sciences, University of Chinese Academy of Sciences, Beijing 100049, China}

\author{Tao Shi}
\email{tshi@itp.ac.cn}
\affiliation{CAS Key Laboratory of Theoretical Physics, Institute of Theoretical Physics, Chinese Academy of Sciences, Beijing 100190, China}
\affiliation{CAS Center for Excellence in Topological Quantum Computation, University of Chinese Academy of Sciences, Beijing 100049, China}

\author{Su Yi}
\email{syi@itp.ac.cn}
\affiliation{CAS Key Laboratory of Theoretical Physics, Institute of Theoretical Physics, Chinese Academy of Sciences, Beijing 100190, China}
\affiliation{School of Physical Sciences, University of Chinese Academy of Sciences, Beijing 100049, China}
\affiliation{CAS Center for Excellence in Topological Quantum Computation, University of Chinese Academy of Sciences, Beijing 100049, China}

\begin{abstract}
We study the collapse dynamics of a Bose-Einstein condensate subjected to a sudden change of the scattering length to a negative value by adopting the self-consistent Gaussian state theory for mixed states. Compared to the Gross-Pitaevskii and the Hartree-Fock-Bogoliubov approaches, both fluctuations and three-body loss are properly treated in our theory. We find a new type of collapse assisted by fluctuations which amplify the attractive interaction between atoms. Moreover, the calculation of the fluctuated atoms, the entropy, and the second-order correlation function showed that the collapsed gas was significantly deviated from a pure state.

\end{abstract}

\maketitle

\section{I\lowercase{ntroduction}}
Although it has been over two decades since the Bosenova experiment~\cite{CllRb1}, accurate theoretical description of the collapse dynamics of Bose-Einstein condensates (BECs)~\cite{CllRb2, CllRb3, 3loss} subjected to a sudden change of interatomic interaction to sufficiently attractive is still elusive. Whereas the theoretical simulations based on the time-dependent Gross-Pitaevskii equation (GPE) with a three-body loss term provides an excellent qualitative understanding to many aspects of the experiments~\cite{Ueda1, GPE2, Ueda2, GPE4, GPE5, GPE6, GPE7,confinment,3BlGPE}, satisfactory quantitative agreement with the experimental observations have not been achieved. These failures may be ascribed to the neglect of high order effects such as excitation or fluctuations driven by the dynamics of the condensates.

In Ref.~~\cite{addfluc1}, Calzetta and Hu considered the impact of the fluctuations on the evolution of the condensates. Similarly, Yurovsky introduced the fluctuations by a linearizing the exact quantum equations of motion for the field operators and attributed loss from the condensate to the growth of the fluctuation~\cite{addfluc2}. Calzetta also showed that the growth of fluctuations led to a shorter collapse time for a collapsing condensate~\cite{addfluc3}. It should be noted that fluctuations were not self-consistently included in the above studies. In an improved treatment, Milstein {\it et al}. studied the collapse dynamics of condensate using the Hartree-Fock-Bogoliubov (HFB) theory~\cite{HFB1}; however, three-body loss was completely ignored in this work. In similar simulations employing the HFB theory, three-body loss was only taken into account in the evolution equation for the condensate~\cite{HFB2,HFB3,HFB4}. Thus the fluctuations are essentially treated as a pure state. The collapses were also simulated using the truncated Wigner method with random noise and a background thermal component in the initial state~\cite{HFB_TWA,HFB3}.

In the present work, we revisit the collapse dynamics of a trapped single-component condensate by using the Gaussian-state theory (GST) for mixed states~\cite{quannon}. By adopting a Gaussian formed density matrix, we derive, from the master equation, the dynamical equations for condensate wave function, the normal and anomalous fluctuations in the presence of three-body loss. These equations combined with the initial state obtained from the Gaussian state calculations provide a self-consistent description for the coherent condensate, excitation, and quantum depletion. Our theory is equivalent to the HFB theory expect that the three-body loss is now properly treated in the dynamic equations for both condensate and fluctuations. To make the numerical simulations tractable, we assume a spherical symmetry for the system regardless of the realistic experimental setups. As a result, the main purpose of this work is not to quantitative reproduce the experimental measurements. Instead, we focus on the new physics originating from the beyond mean field effects. In fact, we identify the {\em deferred collapses} which is assisted by the fluctuations. As a result, the critical interaction strength for the weak collapse is smaller than that predicted by GPE. In addition, we show that due to the atom decay and strong interaction during collapse, a large fraction of atom are transferred into the fluctuations in collapse such that the collapsed gas was significantly deviated from a pure state. It is therefore inappropriate to treat the collapsed atom as a pure coherent state, although the calculation for atom number of the collapsed condensate do not appear to have much difference.

This paper is organized as follows. In Sec.~\ref{secform}, we introduce our model and derive the dynamic equations for the condensate, the normal and the anomalous Green functions based on the master equation. In Sec.~\ref{secstate}, we unveil the structure of the fluctuations by analyzing the normal and the anomalous Green functions. Our simulation results are presented in Sec.~\ref{secresu}. In particular, we show that there exists a new type of collapse assisted by the fluctuations. Finally, we conclude in Sec.~\ref{secconcl}.

\section{F\lowercase{ormulation}}\label{secform}
We consider a condensate of $N$ trapped Bose atoms interacting via $s$-wave collision. In the second-quantized form, the Hamiltonian of the system reads
\begin{align}
H =& \int {\rm d}{\bm r} \hat{\psi}^{\dagger}({\bm r}) \hat h_{0} \hat{\psi}({\bm r}) + \frac{g_{2}}{2} \int {\rm d}{\bm r} \hat{\psi}^{\dagger}({\bm r}) \hat{\psi}^{\dagger}({\bm r}) \hat{\psi}({\bm r}) \hat{\psi}({\bm r}),\label{hamil}
\end{align}
where $\hat{\psi}({\bm r})$ is the field operator, $\hat h_{0}=- \hbar^{2} \nabla^{2} / (2m) +m \omega_{\rm ho}^{2} r^{2} /2$ is the single-particle Hamiltonian with $m$ being the mass of the atom and $\omega_{\rm ho}$ the frequency of the isotropic harmonic trap, $g_{2} = 4 \pi \hbar^{2} a_{s} / (2m)$ is the two-body interaction strength with $a_s$ being the $s$-wave scattering length.

In the presence of atom loss, the system is described by the density matrix $\rho$ satisfying the Lindblad equation
\begin{align}
{\rm i} \hbar \partial_{t} \rho &= \left[H, \rho \right] - {\rm i} \hbar \frac{\gamma_{3}}{3!} \left\{ \int {\rm d}{\bm r}[\psi^{\dagger}({\bm r})]^3 \psi({\bm r})^3, \rho \right\} \nonumber\\
&\quad + {\rm i} \hbar \frac{\gamma_{3}}{3} \int {\rm d}{\bm r}[\psi^{\dagger}({\bm r})]^3 \rho \psi({\bm r})^3 \label{Lindblad},
\end{align}
where $\{\cdot,\cdot\}$ represents the anticommutator, $\hat{\psi}^{3}({\bm r})$ is the jump operator describing the three-body loss with $\gamma_{3}$ being the loss coefficient. Within the framework of GST, the system is described by three order parameters: the coherent state wave function $\phi({\bm r}) = {\rm Tr}\left[\rho \hat{\psi}({\bm r})\right]$, the Green function $G({\bm r}, {\bm r}') = {\rm Tr}\left[ \rho \delta \hat{\psi}^{\dagger}({\bm r}^{\prime}) \delta \hat{\psi}({\bm r}) \right]$, and the anomalous Green function $F({\bm r}, {\bm r}') = {\rm Tr}\left[ \rho \delta \hat{\psi}({\bm r}^{\prime}) \delta \hat{\psi}({\bm r}) \right]$, where $\delta \hat{\psi}({\bm r})=\hat{\psi}({\bm r})-\phi({\bm r})$ is the fluctuation field. Apparently, $G$ and $F$ characterize the fluctuation of the system. To find the dynamical equation of $\phi$, we multiply Eq.~\eqref{Lindblad} by $\hat{\psi}$ and then take the trace, which leads to
\begin{widetext}
\begin{align}
{\rm i} \partial_{t} \phi({\bm r}) & = \hat h_{0} \phi({\bm r}) + g_{2} \left[ \vert \phi({\bm r}) \vert^{2} + 2G({\bm r}, {\bm r}) \right] \phi({\bm r}) + g_{2} F({\bm r}, {\bm r}) \phi^{\ast}({\bm r})-{\rm i} \hbar\frac{\gamma_{3}}{2} \left[ \vert \phi({\bm r}) \vert^{4} \phi({\bm r}) + 6 G({\bm r}, {\bm r}) \vert \phi({\bm r}) \vert^{2} \phi({\bm r})\right. \notag \\
&\quad\left.+ 3 F({\bm r}, {\bm r}) \vert \phi({\bm r}) \vert^{2} \phi^{\ast}({\bm r}) + F^{\ast}({\bm r}, {\bm r}) \phi^{3}({\bm r}) + 6 G^{2}({\bm r}, {\bm r}) \phi({\bm r})+ 3 \vert F({\bm r}, {\bm r}) \vert^{2} \phi({\bm r}) + 6 G({\bm r}, {\bm r}) F({\bm r}, {\bm r}) \phi^{\ast}({\bm r}) \right].\label{tevophi}
\end{align}
Following the similar procedure, we obtain the dynamical equations for $G({\bm r}, {\bm r}')$ and $F({\bm r}, {\bm r}')$ as
\begin{align}
i \partial_{t} G({\bm r}, {\bm r}') &= \int {\rm d}{\bm r}'' \Big\{ \mathcal{E}({\bm r}, {\bm r}'') G({\bm r}'', {\bm r}') + \Delta({\bm r}, {\bm r}'') F^{\ast}({\bm r}'', {\bm r}')- \left[ \mathcal{E}({\bm r}', {\bm r}'') G({\bm r}'', {\bm r}) + \Delta({\bm r}', {\bm r}'') F^{\ast}({\bm r}'', {\bm r}) \right]^{\dagger} \Big\},\label{tevog}\\
i \partial_{t} F({\bm r}, {\bm r}') &= \Delta({\bm r}, {\bm r}') \delta({\bm r} - {\bm r}') + \int {\rm d}{\bm r}'' \Big\{ \mathcal{E}({\bm r}, {\bm r}'') F({\bm r}'', {\bm r}')+ G({\bm r}, {\bm r}'') \Delta({\bm r}'', {\bm r}') \nonumber\\
&\qquad\qquad\qquad\qquad\qquad\qquad\quad+ \big[ \mathcal{E}({\bm r}', {\bm r}'') F({\bm r}'', {\bm r})+ G({\bm r}', {\bm r}'') \Delta({\bm r}'', {\bm r}) \big]^{\rm T} \Big\},\label{tevof}
\end{align}
where
\begin{align}
\mathcal{E}({\bm r}, {\bm r}^{\prime}) &= \hat h_{0} + 2 g_{2} \Big[ \vert \phi({\bm r}) \vert^{2} + G({\bm r}, {\bm r}) \Big]\delta({\bm r}-{\bm r}^{\prime}) \notag \\
&\quad - {\rm i} \hbar\frac{3 \gamma_{3}}{2} \Big[ \vert \phi({\bm r}) \vert^{4} + 4 G({\bm r}, {\bm r}) \vert \phi({\bm r}) \vert^{2} + F({\bm r}, {\bm r}) \phi^{\ast 2}({\bm r})+ F^{\ast}({\bm r}, {\bm r}) \phi^{2}({\bm r}) + 2 G^{2}({\bm r}, {\bm r}) + \vert F({\bm r}, {\bm r}) \vert^{2} \Big]\delta({\bm r}-{\bm r}^{\prime}),  \\
\Delta({\bm r}, {\bm r}^{\prime}) &= g_{2} \Big[ \phi^{2}({\bm r}) + F({\bm r}, {\bm r}) \Big]\delta({\bm r}-{\bm r}^{\prime})\nonumber\\
&\quad -{\rm i} \hbar\gamma_{3} \Big[ \vert \phi({\bm r}) \vert^{2} \phi^{2}({\bm r}) + 3 G({\bm r}, {\bm r}) \phi^{2}({\bm r})+ 3 F({\bm r}, {\bm r}) \vert \phi({\bm r}) \vert^{2} + 3 G({\bm r}, {\bm r}) F({\bm r}, {\bm r}) \Big]\delta({\bm r}-{\bm r}^{\prime}).
\end{align}
\end{widetext}
It can be easily shown that, when the quantum fluctuations $G$ and $F$ are ignorable, Eq.~\eqref{tevophi} reduces to the GPE with three-body loss being included~\cite{3BlGPE}, i.e.,
\begin{align}
{\rm i} \hbar \partial_{t} \phi({\bm r}) & = \left[\hat h_{0} + g_{2} \vert \phi({\bm r}) \vert^{2}-{\rm i} \hbar \frac{\gamma_{3}}{2} \vert \phi({\bm r}) \vert^{4}\right] \phi({\bm r}).\label{gpe}
\end{align}
In addition, Eqs.~\eqref{tevog} and \eqref{tevof} are exact the HFB equations for the normal and anomalous Green functions if the three-body loss is neglected.

Physical quantities can be conveniently expressed in terms of these order parameters. For example, the density of the gas is
\begin{align}
	n({\bm r}) = {\rm Tr}[\rho \hat{\psi}^{\dagger}({\bm r}) \hat{\psi}({\bm r})] = |\phi({\bm r})|^2+ G({\bm r}, {\bm r}),
\end{align}
from which we deduce that the number of atoms in the coherent state and in the fluctuation are $N_C=\int {\rm d}{\bm r}|\phi({\bm r})|^2$ and $N_F=\int {\rm d}{\bm r}G({\bm r},{\bm r})$, respectively.
Moreover, the total energy is $E = {\rm Tr}(\rho H) = E_{\rm kin} + E_{\rm int}$, where
\begin{align}
	E_{\rm kin} &= \int {\rm d}{\bm r} \left[ \phi^{\ast}({\bm r}) h_{0}({\bm r}) \phi({\bm r}) + \lim_{{\bm r}' \rightarrow {\bm r}} h_{0}({\bm r}') G({\bm r}', {\bm r}) \right] \label{engkin}
\end{align}
and
\begin{align}
	E_{\rm int} &= \frac{g_{2}}{2} \int {\rm d}{\bm r} \Big\{ \vert \phi({\bm r}) \vert^{4} + \left[ \phi^{2}({\bm r}) F^{\ast}({\bm r}, {\bm r}) + {\rm c.c.} \right]\nonumber\\
	&\qquad\qquad+ 4 \vert \phi({\bm r}) \vert^{2} G({\bm r}, {\bm r}) + 2 G({\bm r}, {\bm r})^{2} + \vert F({\bm r}, {\bm r}) \vert^{2} \Big\}\label{engint}
\end{align}
are the kinetic and the interaction energies, respectively. Interestingly, in $E_{\rm int}$, there are more terms contributed by the fluctuations through $G$ and $F$, which suggests that the appearance of the fluctuations may amplify the interaction. This observation can be most easily confirmed by considering a macroscopic squeezed vacuum state, for which the attractive interaction is amplified by a factor of three~\cite{Shi2019}.

We shall study the collapse dynamics by numerically evolving Eqs. \eqref{tevophi}-\eqref{tevof} simultaneously. To make our simulations numerically manageable, we utilize the spherical symmetry of the system by assuming that the order parameters are only functions of radii, i.e., $\phi(r)$, $G(r,r')$, and $F(r,r')$. To compare with the GPE theory, we shall also simulate the collapse dynamics using Eq.~\eqref{gpe} by assuming that condensates are described by a pure coherent state.

\section{C\lowercase{haracterization of the fluctuations}}\label{secstate}
Unlike in a pure Gaussian state where the fluctuations always represent the squeezing, fluctuations in a mixed Gaussian state also contain occupations of the quasiparticle states. To analyze the properties of the mixed Gaussian state, let us first write down the density matrix,
\begin{align}
	\rho=\frac{{\rm e}^{-\hat K}}{Z}
\end{align}
where $\hat K$ is a Hermitian operator and partition function $Z={\rm Tr}(\rho)$. In the Nambu basis $\delta\hat{\Psi}({\bm r})=\left(\delta\hat{\psi}({\bm r}), \delta\hat{\psi}^{\dagger}({\bm r})\right)^{\rm T}$, $\hat K$ can be further expressed as
\begin{align}
	\hat K = \frac{1}{2} \int {\rm d}{\bm r}d{\bm r}'\delta \hat{\Psi}^{\dagger}({\bm r}) \Omega({\bm r},{\bm r}') \delta \hat{\Psi}({\bm r}'),
\end{align}
where $\Omega({\bm r},{\bm r}')=\begin{pmatrix}A({\bm r},{\bm r}')&B({\bm r},{\bm r}')\\ [B({\bm r},{\bm r}')]^*&[A({\bm r},{\bm r}')]^*\end{pmatrix}$ subjected to the conditions
\begin{align}
	[A({\bm r}',{\bm r})]^*=A({\bm r},{\bm r}')\mbox{ and }B({\bm r}',{\bm r})=B({\bm r},{\bm r}').\label{omgcond}
\end{align}
Alternatively, $A$ and $B$ can be regarded as matrices with ${\bm r}$ and ${\bm r}'$ being the indices for the matrix elements. As a result, conditions~\eqref{omgcond} simply implies $A^\dagger=A$ and $B^{\rm T}=B$.

To diagonalize $\hat K$, we introduce the Bogoliubov transformation
\begin{align}
	\delta\hat{\Psi}({\bm r})=S({\bm r})\hat{\beta},\label{bogtran}
\end{align}
where $\hat\beta=\begin{pmatrix}\hat {\boldsymbol b}\\\hat {\boldsymbol b}^\dag\end{pmatrix}$ and $S({\bm r}) = \begin{pmatrix}{\boldsymbol u}({\bm r}) & {\boldsymbol v}^{\ast}({\bm r}) \\{\boldsymbol v}({\bm r}) & {\boldsymbol u}^{\ast}({\bm r}) \\\end{pmatrix}$. More specifically, $\hat{\boldsymbol b}=(\hat b_1,\hat b_2,\ldots)^{\rm T}$ are Bogoliubov quasiparticles and ${\boldsymbol u}({\bm r})=(u_1({\bm r}),u_2({\bm r}),\ldots,u_i({\bm r}),\ldots)$ and ${\boldsymbol v}({\bm r})=(v_1({\bm r}),v_2({\bm r}),\ldots,v_i({\bm r}),\ldots)$ are the mode functions. Here we treat ${\boldsymbol u}$ and ${\boldsymbol v}$ as matrices with $i$ and ${\bm r}$ being the (discrete) column and (continuous) row indices, respectively. Since Bogoliubov quasiparticles satisfy the bosonic commutation relations $[\hat\beta,\hat{\beta}^\dag]=\sigma_z\otimes I$, $S$ must be a symplectic matrix, i.e.,
\begin{align}
S({\bm r})(\sigma_z\otimes I) S^\dagger({\bm r}')=\Sigma_z({\bm r}-{\bm r}'),\label{symplet}
\end{align}
where $I$ is an identity matrix and $\Sigma_z({\bm r}-{\bm r}')=\sigma_z\otimes\delta({\bm r}-{\bm r}')$. Writing out this equation explicitly, we obtain the completeness relation for the mode functions
\begin{align}
\sum_i[u_i({\bm r})u_i^*({\bm r}')-v_i^*({\bm r})v_i({\bm r}')]=\delta({\bm r}-{\bm r}').
\end{align}
Moreover, multiplying $\Sigma_z({\bm r}'-{\bm r}'')S({\bm r}'')$ from left to both sides of the Eq.~\eqref{symplet}, we obtain the normalization conditions:
\begin{align}
S^\dagger({\bm r}')\Sigma_z({\bm r}'-{\bm r}'')S({\bm r}'')=\sigma_z\otimes I
\end{align}
or, equivalently,
\begin{align}
\int {\rm d}{\bm r}[u_i({\bm r})u_j^*({\bm r})-v_i({\bm r})v_j^*({\bm r})]&=\delta_{ij}.
\end{align}

To proceed further, we assume that $\Omega$ is symplectically diagonalized by $S$ as
\begin{align}
S^\dagger\Omega S=D,\label{sympdiag}
\end{align}
where $D=I_2\otimes {\boldsymbol d}$ with $I_2$ being a $2\times 2$ identity matrix and ${\boldsymbol d}={\rm diag}\{d_1,d_2,\ldots,d_i,\ldots\}$ a diagonal matrix. Equation~\eqref{sympdiag} can be transformed into the familiar Bogoliubov equation
\begin{align}
\Sigma_z\Omega S=S\Sigma_z D.
\end{align}
In the quasiparticle basis, the density matrix can be expressed as
\begin{align}
\rho=Z^{-1}{\rm e}^{-\hat {\boldsymbol b}^\dagger {\boldsymbol d}\hat {\boldsymbol b}}=Z^{-1}\exp\left({-\sum_i d_i\hat b_i^\dag \hat b_i}\right).
\end{align}
Making use of the explicit expression for the Bogoliubov transformation~\eqref{bogtran}, i.e.,
\begin{align}
\delta\hat{\psi}({\bm r})=\sum_i\left[u_i({\bm r}) \hat b_i+v_i^*({\bm r}) \hat b_i^\dag\right],
\end{align}
the normal and anomalous Green functions can be decomposed into the forms $G({\bm r}, {\bm r}') = G_{T}({\bm r}, {\bm r}') + G_{Q}({\bm r}, {\bm r}')$ and $F({\bm r}, {\bm r}') = F_{T}({\bm r}, {\bm r}') + F_{Q}({\bm r}, {\bm r}')$. More specifically,
\begin{align}
G_{T}({\bm r}, {\bm r}')&=\sum_{i} f_i \Big( u^{\ast}_{i}({\bm r}')u_{i}({\bm r}) + v_{i}({\bm r}')v^{\ast}_{i}({\bm r}) \Big),\\
F_{T}({\bm r}, {\bm r}')&=\sum_{i} f_{i}\Big( v^{\ast}_{i}({\bm r}') u_{i}({\bm r}) + u_{i}({\bm r}') v^{\ast}_{i}({\bm r}) \Big),
\end{align}
where $f_{i}={\rm tr}(\rho\hat{b}_i^\dagger\hat{b}_i)=1/({\rm e}^{d_i}-1)$ is the average quasiparticle occupation number on the $i$th mode, in analogy to the thermal occupation number at finite temperature. Therefore, we may say that $G_T$ and $F_T$ characterize the thermal fluctuation even if the temperature of the system is zero. On the other hand,
\begin{align}
G_Q({\bm r}, {\bm r}') &= \sum_{i} v_{i}({\bm r}') v^{\ast}_{i}({\bm r})\nonumber\\
&={\sum_{\alpha=1}^{\infty }N_{S,\alpha}\bar{\phi}_{S,\alpha}({\bm r})\bar{\phi}_{S,\alpha}^{\ast }({\bm r}^{\prime })}\label{gsqu}
\end{align}
and
\begin{align}
F_Q({\bm r}, {\bm r}') &= \sum_{i} \frac{1}{2} \Big( v^{\ast}_{i}({\bm r}') u_{i}({\bm r}) + u_{i}({\bm r}') v^{\ast}_{i}({\bm r}) \Big)\nonumber\\
&={\sum_{\alpha=1}^{\infty }\sqrt{N_{S,\alpha}(N_{S,\alpha}+1)}\bar{\phi}_{S,\alpha}({\bm r})\bar{\phi}_{S,\alpha}({\bm r}^{\prime })}\label{fsqu}
\end{align}
are quantum fluctuation (or quantum depletion) which does not represent actual occupation of the Bogoliubov excitation modes. Moreover, as shown in the second lines of Eqs.~\eqref{gsqu} and \eqref{fsqu}, $G_Q$ and $F_Q$ can be simultaneously diagonalized by a set of orthonormal modes $\{\bar\phi_{S,\alpha}({\bm r})\}$ satisfying $\int {\rm d}{\bm r}\bar{\phi}_{S,\alpha}^*({\bm r})\bar{\phi}_{S,\alpha'}({\bm r})=\delta_{\alpha\alpha'}$. Therefore, similar to those in a pure Gaussian state, $G_Q$ and $F_Q$ characterize squeezing with $N_{S,\alpha}$ being the occupation number in the $\alpha$th squeezed mode $\bar\phi_{S,\alpha}$. Then $N_S=\sum_jN_{S,\alpha}$ is the total number of squeezed atoms. Without loss of generality, we assume that $N_{S,\alpha}$ are sorted in descending order with respect to the index $\alpha$. Thus $\bar{\phi}_{S,1}$ represents the squeezed mode with highest occupation. Interestingly, the condensate is in a macroscopic squeezed state when $\bar{\phi}_{S,1}$ is macroscopically occupied~\cite{Shi2019,Wang2020,Pan2021}. And for weakly attractive condensate, a condensate can even be in a pure single-mode squeezed state with $\bar{\phi}_{S,1}\simeq N$. In this case, it can be clearly seen from Eq.~\eqref{engint} that the interaction energy is amplified by a factor of three~\cite{Shi2019}.

To distinguish different states, it is helpful to compute the second-order correlation function
\begin{align}
g^{(2)}({\bm r}, {\bm r}) &= \frac{{\rm tr}[\rho \hat{\psi}^{\dagger}({\bm r}) \hat{\psi}^{\dagger}({\bm r}) \hat{\psi}({\bm r}) \hat{\psi}({\bm r})]}{{\rm tr}[\rho \hat{\psi}^{\dagger}({\bm r}) \hat{\psi}({\bm r})]^2} \nonumber \\
&= n({\boldsymbol r})^{-2}\Big\{\vert \phi({\bm r}) \vert^{4} + 2 G^{2}({\bm r}, {\bm r}) + \vert F({\bm r}, {\bm r}) \vert^{2}\nonumber\\
&\quad + 4 G({\bm r}, {\bm r}) \vert \phi({\bm r}) \vert^{2}  + 2{\rm Re} [F^{*}({\bm r}, {\bm r}) \phi^{2}({\bm r})]\Big\}.
\end{align}
The following special cases are of particular importance. i) For a pure coherent state, $G$ and $F$ vanish, which leads to $g_{\rm coherent}^{(2)}({\bm r},{\bm r})=1$; ii) For a thermal state, $\phi$, $G_S$, and $F_S$ are all zero. As a result, all $v_i({\bm r})$'s and, subsequently, $F$ vanishes, which further yields $
g_{\rm thermal}^{(2)}({\bm r}, {\bm r}) = 2$; iii) For a pure squeezed state, we have $N_{S,1}\simeq N$. As a result, $\phi$, $G_T$, and $F_T$ vanish, which implies $F({\bm r}, {\bm r}) \approx G({\bm r}, {\bm r})$ and, subsequently, $g_{\rm squeeze}^{(2)}({\bm r}, {\bm r}) \approx 3$. Therefore, measuring $g^{(2)}$ should allow us to identify the state of a collapsed condensate.

Finally, we shall also use the entropy
\begin{align}
\mathcal{S}(\rho) &= -{\rm tr} (\rho \ln\rho)\nonumber\\
&= \sum_{i} \left[ (f_{i} + 1) \ln{(f_{i} + 1)} - f_{i} \ln f_{i} \right]
\end{align}
to measure the deviation of the collapsed condensate from a pure state.

\section{R\lowercase{esults}}\label{secresu}
To systematically explore the collapse dynamics, we first recall that the system is completely specified by the following parameters: atom number $N$, trap frequency $\omega_{\rm ho}$, scattering length $a_s$, and three-body loss coefficient $\gamma_3$. Without loss of generality, the trap frequency is fixed at $\omega_{\rm ho}=(2\pi)\,12.8\,{\rm Hz}$ which is the geometric average of the trap frequencies in three Cartesian directions of the experiment~\cite{CllRb2}. In all simulations, we prepare an initial pure state by numerically solving the imaginary-time equations of motion for a Gaussian state~\cite{Shi2019,Wang2020,Pan2021} under the initial atom number $N(0)$ and scattering length $a_s=a_{\rm init}$ ($\geq0$). We then quench the scattering length to $a_s=a_{\rm final}$ ($<0$) at $t=0$. It should be noted that a trapped BEC with attractive interactions becomes unstable only when the dimensionless parameter (DIP)
\begin{align}
k = \frac{N \vert a_{s} \vert}{a_{\rm ho}}
\end{align}
exceeds a critical value, say $k_{\rm cri}$, where $a_{\rm ho}=\sqrt{\hbar/(m\omega_{\rm ho})}$ is the harmonic oscillator length. For the chosen parameter, we have $a_{\rm ho}=5.77\times 10^4a_B$ with $a_B$ being the Bohr radius. There exist many studies on the critical interaction strength of a trapped condensate~\cite{attractiveNLSE, stableLi,Tunneling1, Tunneling2,NcUeda,collapsingDynamic}. The dynamics of the condensate is then simulated by numerically evolving Eqs. \eqref{tevophi}-\eqref{tevof}. We point out that, to minimize the impact of $a_{\rm init}$ on $k_{\rm cri}$, it is preferable to choose $a_{\rm init}=0$. However, in order to obtain an initial state with nonvanishing fluctuations, we normally adopt a very small $a_{\rm init}$ in our simulations.

\begin{figure}[ptb]
\centering
\includegraphics[width=0.9\columnwidth]{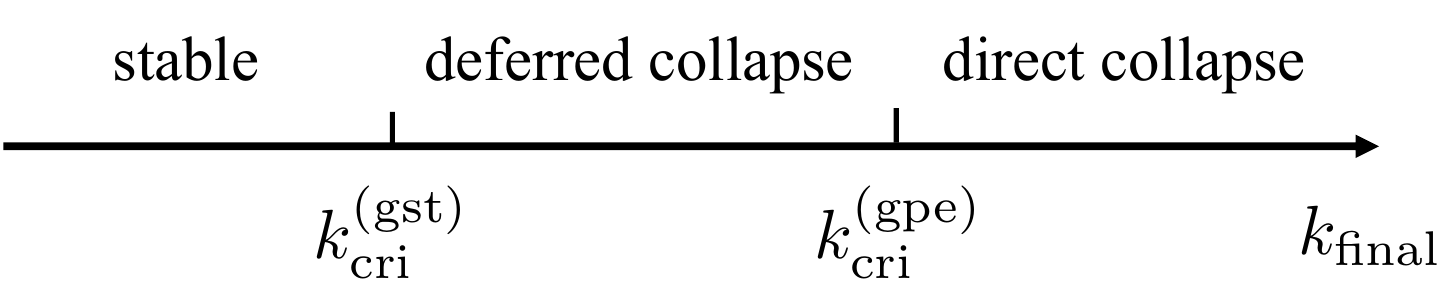}
\caption{(color online). Schematic plot for the collapse types on the axis of DIP.}
\label{phase}
\end{figure}

As shall be shown, for a same set of $N(0)$, $a_{\rm init}$, and $\gamma_3$, the GST and GPE approaches may lead to two distinct critical interaction strengths, say $k_{\rm cri}^{\rm (gst)}$ and $k_{\rm cri}^{\rm (gpe)}$, which satisfy $k_{\rm cri}^{\rm (gst)}<k_{\rm cri}^{\rm (gpe)}$. Consequently, based on the final interaction parameter $k_{\rm final}$, we categorize the collapses into i) the direct collapse that happens when $k_{\rm final}>k_{\rm cri}^{\rm (gpe)}$ and ii) the deferred collapse which is stimulated by the fluctuations and occurs under the condition $k_{\rm cri}^{\rm (gpe)}>k_{\rm final}>k_{\rm cri}^{\rm (gst)}$. In other words, a direct collapse also happens in the GPE simulations; while a deferred collapse only occurs when we simulate it using GST. In Fig.~\ref{phase}, we schematically show the parameter regimes for different types of collapses.

\subsection{Direct collapses}

\begin{figure}[ptb]
\centering
\includegraphics[width=0.95\columnwidth]{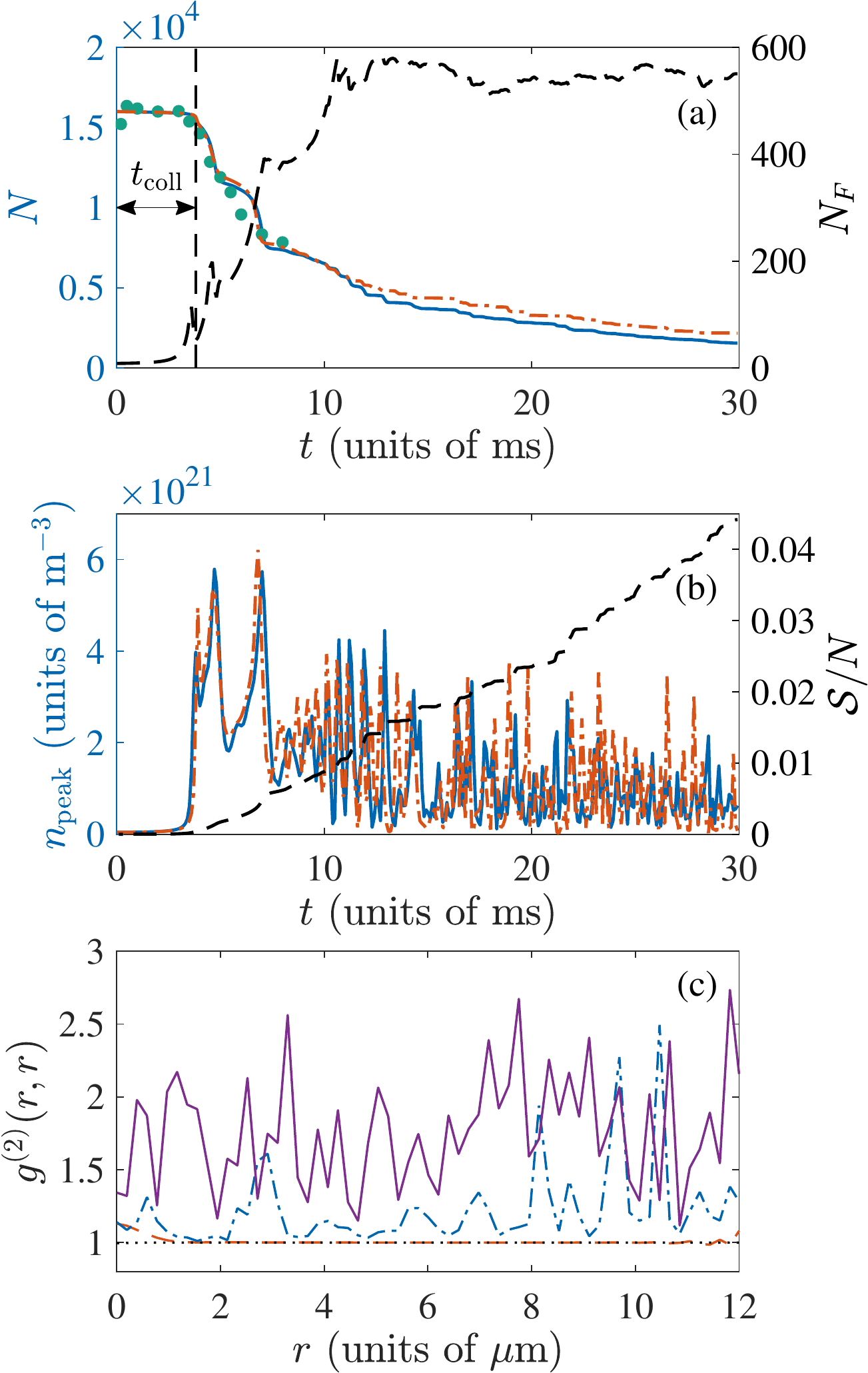}
\caption{(color online). (a) Time dependence of the total atom number computed via GST (solid line) and GPE (dash-dotted line). Filled circles ($\bullet$) represent the experimental data~\cite{CllRb2}. The black dashed line shows the time dependence of $N_F$ (right $y$ axis) obtained using GST. (b) Time dependence of the peak density computed via GST (solid line) and GPE (dash-dotted line). The black solid line shows the time dependence of the entropy per atom (right $y$ axis). (c) Second-order correlation function $g^{(2)}(r,r)$ for $t=0$ (dotted line), $3$ (dashed line), $10$ (dash-dotted line), and $30\,{\rm ms}$ (solid line). Other parameters are $N(0)=1.6\times 10^4$, $a_{\rm init}=7a_B$, $a_{\rm final}=-30 a_{B}$, and $\gamma_{3} = 3 \times 10^{-27} {\rm cm}^{6} / {\rm s}$.}
	\label{cg3loss2}
\end{figure}

As an example for direct collapses, we perform simulations with the same set of control parameters as those used in the experiment~\cite{CllRb2}, i.e., $N(0)=1.6\times 10^4$, $a_{\rm init}=7a_B$, and $a_{\rm final}=-30 a_{B}$. Correspondingly, the DIP $k_{\rm final}=8.32$ is much larger than the critical interaction strength. Figure~\ref{cg3loss2}(a) and (b) plot the time dependence of the total atom number $N$ and the peak condensate density $n_{\rm peak}$, respectively. Here the three-body loss coefficient is taken as $\gamma_{3} = 3 \times 10^{-27} {\rm cm}^{6} / {\rm s}$, a value obtained by fitting the atom number with experiment data~\cite{CllRb2} [dots in Fig.~\ref{cg3loss2}(a)]. For comparison, we also present the results from the GPE simulation (dash-dotted lines). As can be seen, for atom number $N(t)$, the results obtained via both approaches are in good agreement. However, for the peak density, a large discrepancy appears when $t$ is roughly larger than $7\,{\rm ms}$. In addition, our results are in qualitative agreement with the simulations presented in Refs.~\cite{Ueda1,Ueda2}.

For a typical direct collapse, after the scattering length is quenched, $N$ roughly remains constant for some time and then experiences a sudden decay which signals a collapse of the condensate. The time of this collapse defines the collapse time $t_{\rm coll}$. After $t_{\rm coll}$, collapses occur intermittently such that $N(t)$ decays stepwise. Associated with each collapse, there is a spike on the $n_{\rm peak}$-$t$ curve, indicating that the condensate first implodes and then explodes. The underlying reason for the formation of the spikes was previously studied in Refs.~\cite{Ueda1,Ueda2}. Specifically, during an implosion, condensate shrinks and its peak density abruptly increases. Consequently, both the kinetic and the interaction energies increase. This process is also accompanied by the increase of the three-body loss which lowers $n_{\rm peak}$. When the atom loss rate becomes larger than accumulation rate of the atoms, the peak density ceases to increase (see below for a detailed analysis). Now, because the kinetic and interaction energies are proportional to $n_{\rm peak}$ and $n_{\rm peak}^2$, respectively, the attractive interaction energy decreases faster than the kinetic energy. As a result, the attraction is insufficient to bound gas such that the condensate starts to explode and the peak density is quickly lowered.

This observation can be understood by a simple model described below. Within $t_{\rm coll}$, the squeezed atoms in condensate is negligible such that the condensate is solely described by $\phi(r)$. In addition, as the shape of the condensate is well maintained, $\phi$ can then be approximated by a Gaussian function
\begin{align}
\phi(r)=\left[\frac{N}{\pi^{3/2} \sigma(t)^{3}}\right]^{1/2} {\rm e}^{-r^2/[2\sigma(t)^2]-{\rm i}r^2\beta(t)},\label{gauans}
\end{align}
where $\sigma$ is the width of the condensate and $\beta(t)$ accounts for the dynamics due to the kinetic energy. It can be shown that $\sigma$ satisfies the dynamics equation
\begin{align}
m\frac{d^2\sigma}{dt^2}&=-\frac{\partial V_{\rm eff}(\sigma)}{\partial \sigma},\label{effdyn}
\end{align}
where
\begin{align}
V_{\rm eff}(\sigma)=\frac{1}{2}\hbar\omega_{\rm ho}\left(\frac{\sigma^2}{a_{\rm ho}^2} + \frac{a_{\rm ho}^2}{\sigma^2} - \frac{4k_{\rm final}}{3\sqrt{2\pi}} \frac{a_{\rm ho}^3}{\sigma^3}\right), \label{veff}
\end{align}
is the effective potential experienced by a particle with mass $m$. Clearly, $V_{\rm eff}$ contains the contributions from potential, kinetic, and interaction energies. Once $\sigma(t)$ is obtained, $\beta(t)$ can be evaluated according to
\begin{align}
\beta(t) = \frac{m}{2\hbar\sigma}\frac{{\rm d}\sigma}{{\rm d}t}.\label{effdyn2}
\end{align}

We now use this simple variational wave function to estimate the height of the first spikes on $n_{\rm peak}$-$t$ curve. To this end, we first derive, from Eq.~\eqref{gpe}, a continuity equation
\begin{align}
\partial_t|\phi|^2=-\nabla\cdot{\bm J}-\hbar\gamma_3|\phi|^6,
\end{align}
where ${\bm J}=\frac{\hbar}{m}{\rm Im}(\phi^*\nabla\phi)$. Making use of the ansatz~\eqref{gauans}, the continuity equation reduces to
\begin{align}
\frac{{\rm d}}{{\rm d}t} n_{\rm peak}(t) = \frac{6\hbar\beta(t)}{m}  n_{\rm peak}(t) - \gamma_{3} n^{3}_{\rm peak}(t),\label{dnpeakdt}
\end{align}
where
\begin{align}
n_{\rm peak}(t) = \frac{N_{c}}{\pi \sqrt{\pi} \sigma(t)^{3}}\nonumber
\end{align}
is the peak density of the Gaussian density profile. The time for the first spike, i.e., $t_{\rm spike}$, can be determined using the condition that the peak density stops growing at $t=t_{\rm spike}$. Then from Eq.~\eqref{dnpeakdt}, we obtain
\begin{align}
\beta(t_{\rm spike}) = \frac{m \gamma_{3}}{6 \hbar} n^{2}_{\rm peak}(t_{\rm spike}).\label{btnt}
\end{align}
Now, to determine $t_{\rm spike}$, we numerically solve Eqs.~\eqref{effdyn} and \eqref{effdyn2} such that the condition~\eqref{btnt} is satisfied. With the parameters used in Fig.~\ref{cg3loss2}, we find that $t_{\rm spike}\approx 4\,{\rm ms}$ and $n_{p}(t_{\rm spike})\approx1.0637 \times 10^{21}\, {\rm m}^{-3}$, which are in rough agreement with the full numerical simulation.

Next, we compare the GST and the GPE descriptions of the collapse dynamics. To this end, we also plot, in Fig.~\ref{cg3loss2}(a), the number of the fluctuated atoms $N_F$ as a function of time $t$. Immediately after the collapse starts, $N_F$ quickly increases and then saturates at about $550$ atoms after $t\approx10\,{\rm ms}$. In particular, $N_F/N$ can be as large as $30\%$ at $t=30\,{\rm ms}$, which suggests that the statistical property of the condensate might be dramatically modified. Moreover, as shown in Fig.~\ref{cg3loss2}(b), the entropy of the system monotonically increases and becomes nearly saturated at large $t$,  indicating that the system is significantly deviated from a pure state. To gain more details, we present, in Fig.~\ref{cg3loss2}(c), the second-order correlation function $g^{(2)}(r,r)$ at various times. As expected, the second-order correlation function is unity for the initial state. Then for $t=3\,{\rm ms}$, $g^{(2)}(r,r)$ begins to deviate from unity at the high-density region where the three-body loss is important. Finally, at later times, $g^{(2)}(r,r)$ is significantly deviated from unity along the whole radial direction. These results suggest that the fluctuations should be taken into account for an accurate description of the collapse dynamics.

\begin{figure}[ptb]
\includegraphics[width=0.95\columnwidth]{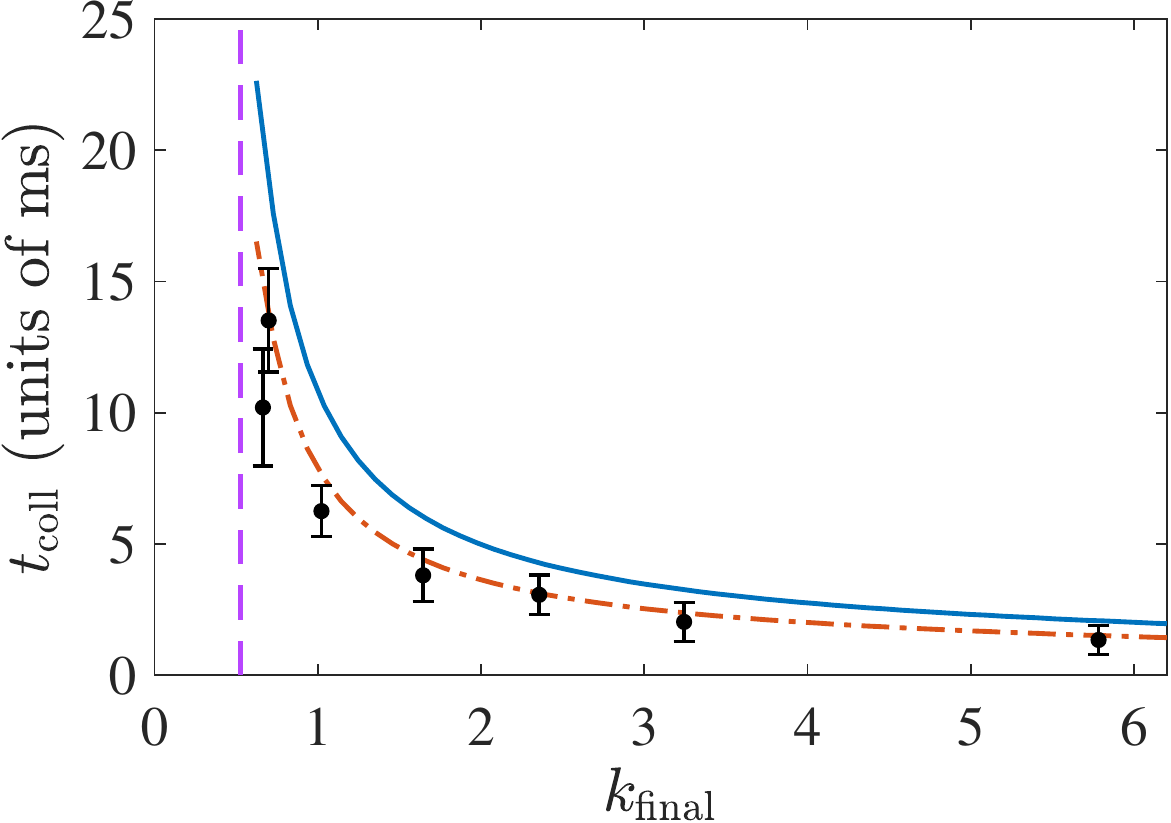}
\caption{(color online). $t_{\rm coll}$ versus $k_{\rm final}$ for $N(0)=6000$, $a_{\rm init}=10^{-4} a_B$, and $\gamma_{3} = 3 \times 10^{-27} {\rm cm}^{6} / {\rm s}$. The filled circle ($\bullet$) are the experimental data extracted from Ref.~\cite{CllRb2}, the solid line represents the GST result, and the dash-dotted line is the GST result multiplied by a factor $12.8/17.5$. The vertical dashed line marks $k_{\rm cri}$.}\label{tcoll1}
\end{figure}

To gain more insight into the collapse dynamics, we explore how the collapse time depends on $k_{\rm final}$. In Fig.~\ref{tcoll1}, we plot the numerically computed $t_{\rm coll}$ as a function of $k_{\rm final}$ for $N(0)=6000$, $a_{\rm init}=10^{-4}\,a_B$, and $\gamma_{3} = 3 \times 10^{-27} {\rm cm}^{6} / {\rm s}$. Interestingly, unlike the computation of the ground state, we also find that collapses occur even when $k_{\rm final}\approx 0.52$, in agreement with the result in Ref.~\cite{collapsingDynamic}. However, there exists a systematical discrepancy between the numerical and the experimental results, originating from the distinct trap frequency used in the simulations. In fact, the collapse time is closely related to the trap frequency as, after the scattering length is quenched, all atoms accumulate at the trap center at roughly $t=T_{\rm ho}/4$ ($T_{\rm ho}\equiv 2\pi/\omega_{\rm ho}$) such that the highe    st density (where the collapse most likely occurs) is achieved~\cite{frequency}. For an anisotropic trap as that used in experiment, this time is determined by the radial trap frequency $(2\pi)17.5\,{\rm Hz}$~\cite{CllRb1,CllRb2} which is larger than the trap frequency along the axial direction. Therefore, to compare with the experiment, we rescale our numerical results by the factor $12.8/17.5$, which, as shown in Fig.~\ref{tcoll1} by the dash-dotted line, leads to a better agreement.

\subsection{Fluctuation assisted deferred collapses}

\begin{figure}[ptb]
\includegraphics[width=0.9\columnwidth]{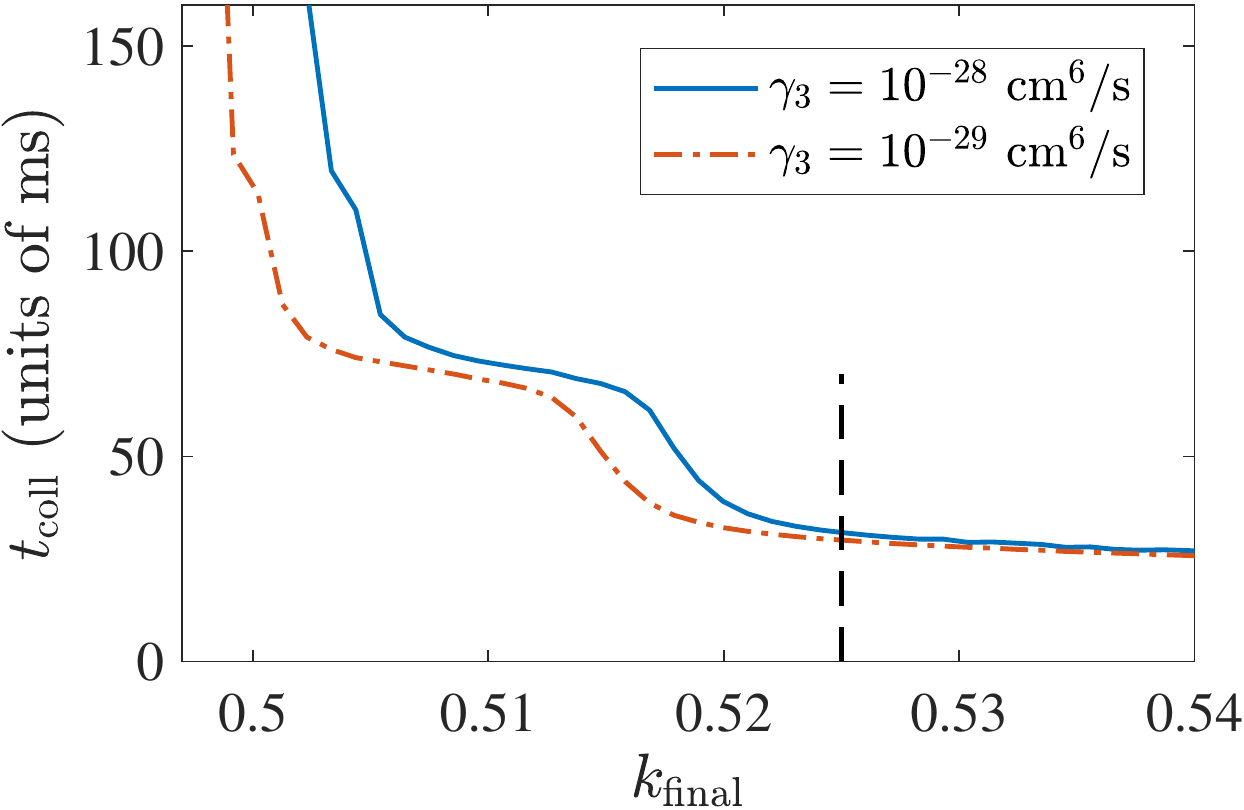}
	\caption{(color online). $t_{\rm coll}$ as a function of $k_{\rm final}$ for $\gamma_{3} = 10^{-28} {\rm cm}^{6} / {\rm s}$ (solid line) and $10^{-28} {\rm cm}^{6} / {\rm s}$ (dash-dotted line). Other parameters are $N(0)=6000$ and $a_{\rm init}=10^{-4}\,a_B$. The black dashed line mark the critical DIP obtained with GPE.}
\label{cllt}
\end{figure}

In order to observe deferred collapses, we have to reduce the value of the three-body loss coefficient; otherwise, the atom number may decay too fast such that $k_{\rm final}$ is significantly lowered and the collapse is suppressed. In Fig.~\ref{cllt}, we plot the collapse time as a function of $k_{\rm final}$ for $N(0)=6000$, $a_{\rm init}=10^{-4}\,a_B$, and $\gamma_{3} = 10^{-28}$ and $10^{-29} {\rm cm}^{6} / {\rm s}$. As can be seen, although $k_{\rm cri}^{\rm (gpe)}\approx 0.52$ are roughly the same in both cases, $k_{\rm cri}^{\rm (gst)}$ are now $0.505$ and $0.495$ for $\gamma_{3} = 10^{-28}$ and $10^{-29} {\rm cm}^{6} / {\rm s}$, respectively. In addition, it is seen that $t_{\rm coll}$ increases stepwise as $k_{\rm final}$ gradually decreases. We point out that the three-body loss coefficient used here was also used in the earlier theoretical simulations~\cite{Ueda2} and is accessible in realistic experimental systems~\cite{roberts2000magnetic,3loss}.

\begin{figure}[ptb]
\includegraphics[width=0.97\columnwidth]{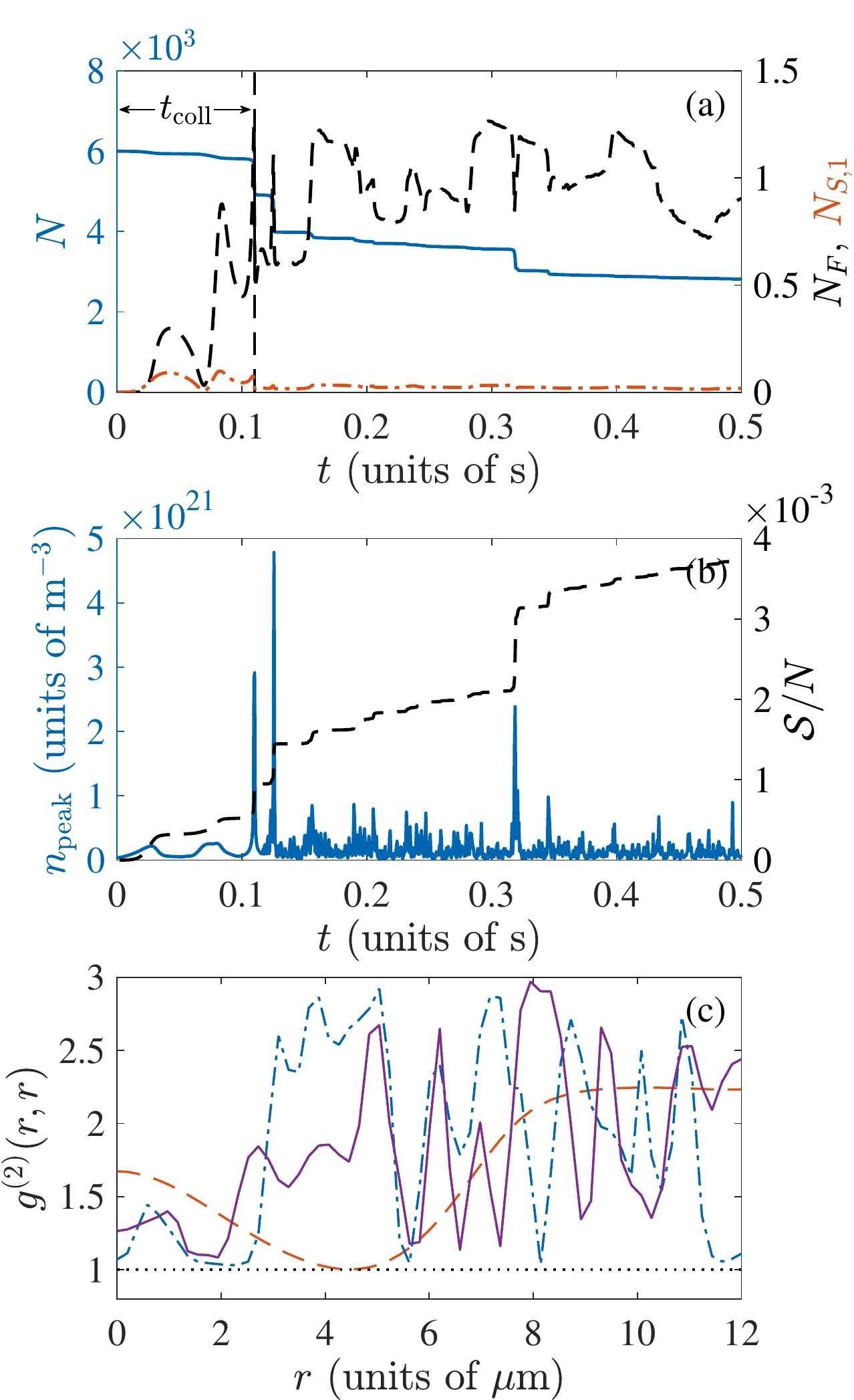}
\caption{(color online). (a) $N(t)$ (solid line), $N_F(t)$ (dashed line), and $N_{S,1}(t)$ (dash-dotted line). (b) Time dependence of the peak density (solid line) and the entropy per atom (dashed line). (c) The second-order correlation function $g^{(2)}(r,r)$ for $t=0$ (dotted line), $0.1$ (dashed line), $0.2$ (dash-dotted line), $0.4\,{\rm s}$ (solid line). The parameters used here are $N(0) = 6000$, $a_{\rm init} = 10^{-4}\,a_B$, $a_{\rm final}=-4.85\,a_B$, and $\gamma_3=10^{-29} {\rm cm}^{6} / {\rm s}$. Correspondingly, the DIP is $k_{\rm final} = 0.5$.}
\label{coparision}
\end{figure}

To proceed further, we plot, in Fig.~\ref{coparision}(a), $N(t)$, $N_F(t)$, and $N_{Q}(t)$ for a typical deferred collapse with $N(0) = 6000$, $a_{\rm init} = 10^{-4}\,a_B$, $a_{\rm final}=-4.85\,a_B$ ($k_{\rm final} = 0.5$), and $\gamma_{3} = 10^{-29} {\rm cm}^{6} / {\rm s}$. Correspondingly, Fig.~\ref{coparision}(b) plots the time dependence of the peak condensate density $n_{\rm peak}$ and the entropy $\mathcal{S}$. As can be seen, once the collapse starts at around $t\approx0.113\,{\rm s}$, the dynamics behavior of the system becomes very similar to that in a direct collapse. Therefore, the feature that differs from a direct collapse lies at its dynamic behavior prior to the collapse. Particularly, as shown in Fig.~\ref{coparision}(b), the peak density $n_{\rm peak}$ oscillates for about $5/2$ periods before collapse. This oscillation corresponds to the breathing mode of the condensate and can be explained using the dynamical equation Eq.~\eqref{effdyn}. In fact, for $k_{\rm final}<0.67$, there exists a local minimum in the effective potential $V_{\rm eff}$. Thus after the scattering length is quenched, $\sigma$ starts to oscillate around the equilibrium width. The oscillation frequency can be analytically obtained by linearizing Eq.~\eqref{effdyn}, which gives rise to the period of the breathing mode
\begin{align}
T_{\rm breathing}=T_{\rm ho} \left(\frac{1}{a^{2}_{\rm ho}} + \frac{3 a^{2}_{\rm ho}}{\sigma_0^{4}} - 8 \frac{k_{\rm final}}{\sqrt{2 \pi}} \frac{a^{2}_{\rm ho}}{\sigma_0^{5}}\right)^{-1/2}.\label{breathmd}
\end{align}
For parameters used in Fig.~\ref{coparision}, Eq.~\eqref{breathmd} yields $T_{\rm breathing}\approx42\,{\rm ms}$ which is in good agreement with numerical simulations. Following this analysis, because the density of the condensate attains the highest value at times that are odd multiples of $T_{\rm breathing}/2$, the $t_{\rm coll}$-$k_{\rm final}$ curve (Fig.~\ref{cllt}) is naturally of the stepwise shape. Accompanying the density oscillation of the condensate, the number of fluctuated atoms also oscillates. In particular, at $t\approx 80\,{\rm ms}$, $N_F$ can be as large as $1000$ and it becomes even larger close the $t_{\rm coll}$. These fluctuated atoms originate from two mechanisms, i.e., the decay induced decoherence and the attractive interaction induced squeezing. Because, as shown in Eq.~\eqref{engint}, the fluctuated atoms amplify the attractive interaction~\cite{Shi2019}, collapse can then be induced when the number of atoms in the fluctuations becomes sufficiently large. As shown in Fig.~\ref{coparision}(a), it should be noted that, among the fluctuated atoms, there is only a small fraction of atoms in the pure squeezed state (quantum depletion). Finally, once the collapse is initiated, the dynamical behavior of the gas, as shown in Fig.~\ref{coparision}(b) and (c) for $n_{\rm peak}$, $\mathcal{S}$, and $g^{(2)}(r,r)$, is very similar to that of a strong collapse, which again suggests that fluctuations should be considered for the studying of the collapse dynamics.

We would also like to point out that deferred collapse found here is stimulated by the fluctuations which is completely different from the delayed collapse previously predicted by Biasi {\it et al}.~\cite{delayedcoll}. Their study was based on the GPE with atom decay mechanism being completely ignored. In addition, the delayed collapses are induced by changing the shape of the initial condensates.

\section{C\lowercase{onclusion and discussion}}\label{secconcl}
In conclusion, we have studied the collapse dynamics of a Bose-Einstein condensate using GST. Compared to the coherent-state-based GPE approach, fluctuations are properly treated at the mean-field level. It has been shown that the presence of the fluctuations leads to a critical interaction strength that is slightly smaller than that predicted by GPE. Moreover, the calculation of the fluctuated atoms, the entropy, and the second-order correlation function showed that the collapsed gas was significantly deviated from a pure state. It is therefore inappropriate to treat the collapsed atom as a pure coherent state, although the calculation for atom number of the collapsed condensate do not appear to have much difference. As our future works, we shall revisit the $d$-wave collapse of dipolar condensates~\cite{dwavecollapse} and study the dynamical formation of quantum droplets in both dipolar and binary condensates~\cite{singledropDy,Cabrera}.

\begin{acknowledgments}
This work was supported by the National Key Research and Development Program of China (Grant No. 2021YFA0718304), by the NSFC (Grants No. 12135018 and No. 12047503), and by the Strategic Priority Research Program of CAS (Grant No. XDB28000000).
\end{acknowledgments}

\bibliography{ref_collapse.bib}

\end{document}